\documentclass[aps,prd,superscriptaddress,notitlepage,showpacs]{revtex4-1}

\usepackage{graphicx}
\usepackage{amsmath}
\usepackage{amssymb}
\usepackage{color}
\usepackage{bm}

\begin{document}

\title{Chiral asymmetry in cold QED plasma in a strong magnetic field}
\date{\today}

\author{Lifang Xia}
\affiliation{Department of Physics, Arizona State University, Tempe, Arizona 85287, USA}

\author{E. V. Gorbar}
\affiliation{Department of Physics, Taras Shevchenko National University of Kiev, Kiev, 03680, Ukraine}
\affiliation{Bogolyubov Institute for Theoretical Physics, Kiev, 03680, Ukraine}

\author{V. A. Miransky}
\affiliation{Department of Applied Mathematics, Western University, London, Ontario N6A 5B7, Canada}

\author{I. A. Shovkovy}
\affiliation{College of Letters and Sciences, Arizona State University, Mesa, Arizona 85212, USA}

\begin{abstract}
The interaction induced chiral asymmetry is calculated in cold QED plasma beyond the weak-field 
approximation. By making use of the recently developed Landau-level representation for the 
fermion self-energy, the chiral shift and the parity-even chiral chemical potential function are 
obtained with the help of numerical methods. The results are used to quantify the chiral 
asymmetry of the Fermi surface in dense QED matter. Because of the weakness of the QED
interactions, the value of the asymmetry appears to be rather small even in the strongest magnetic 
fields and at the highest stellar densities. However, the analogous asymmetry can be substantial 
in the case of dense quark matter.
\end{abstract}


\maketitle

\section{Introduction}
\label{Sec:Intro}

Nowadays the studies of chiral asymmetry in magnetized relativistic matter has drawn attention of 
researchers across diverse areas in physics. Heavy-ion collisions \cite{Kharzeev1,Kharzeev2,Liao}, 
compact stars \cite{Charbonneau,Ohnishi}, the early Universe \cite{Giovannini:1997eg,Boyarsky:2011uy,Tashiro:2012mf}, 
and Dirac/Weyl semimetals \cite{Turner,Vafek} are the main physical systems where such studies are 
relevant. The principal role in generating a chiral asymmetry in relativistic matter with usual vectorlike 
gauge interactions is played by an external magnetic field $\mathbf{B}$. In fact, it is 
the lowest Landau level (LLL) that is primarily responsible for the generation of chiral asymmetry 
in magnetized relativistic matter. The LLL is special because it has fermion spins completely 
polarized: they are directed along the magnetic field for a positive charge and opposite to the
field for a negative one. For massless or ultrarelativistic particles, it is more 
appropriate to talk about their helicity rather than spin. Since the helicity of massless particles 
is the same as their chirality and the chiralities of the left- and right-handed particles are 
opposite, it is easy to understand how a nondissipative axial current 
$\mathbf{j}_5=e\mathbf{B}\mu/(2\pi^2)$ is generated in magnetized relativistic 
fermion matter at nonzero chemical potential $\mu$ \cite{Vilenkin}. This result is 
known in the literature as the chiral separation effect (CSE) (see, e.g., Ref.~\cite{Fukushima}) 
and only the lowest Landau level contributes to the axial current in the free theory 
\cite{Zhitnitsky}.

The chiral magnetic effect (CME) \cite{Kharzeev:2007jp,Fukushima:2008xe} is in a sense a dual 
phenomenon to the chiral separation effect. The CME implies that the chiral asymmetry
in magnetized relativistic matter, e.g., described by a nonvanishing chiral chemical potential 
$\mu_5 \ne 0$, causes a nondissipative electric current $\mathbf{j}=e^2\mathbf{B}\mu_5/(2\pi^2)$.
(See, however, the recent holographic study in Ref.~\cite{Jimenez-Alba:2014iia}, which 
points some fundamental differences between the realization of the CME and CSE.)
Moreover, an interplay of the chiral separation and chiral magnetic effects gives rise to a novel 
type of collective gapless excitations: the chiral magnetic wave (CMW) \cite{CMW1}. 
Indeed, according to the CSE, a local fluctuation of the electric charge density induces a local 
fluctuation of the axial current. The resulting fluctuation of the chiral chemical potential produces 
a local fluctuation of the electric current via the CME. The latter in turn leads again to a local 
fluctuation of electric charge density and thus provides a self-sustaining mechanism for the 
propagation of a chiral magnetic wave. In heavy-ion collisions, such a wave leads to the 
quadrupole CME \cite{chiral-shift-2,CMW2}. One of the observable implications 
of the latter is the splitting of the elliptic flows of positively and negatively charged pions, i.e., 
$v^{\pi^-}_2-v^{\pi^+}_2=r_eA$, where $A$ is the net charge asymmetry 
$A=(\bar{N}^{+}-\bar{N}^{-})/(\bar{N}^{+}+\bar{N}^{-})$ and $r_e$ is the slope parameter 
\cite{CMW2}. Such a splitting was observed by the STAR collaboration 
\cite{Adamczyk:2013kcb,Wang:2012qs,Ke:2012qb} and appears to be 
in agreement with theoretical predictions.

Apart from the CSE and CME, there also exist other related anomalous transport phenomena, e.g., 
the chiral vortical effect (CVE) \cite{CVE1,CVE2,CVE3,CVE4}, the chiral electric effect (CEE) 
\cite{Neiman}, and the chiral charge generation effect (CCGE) \cite{Minwalla,Melgar}. In the free 
theory, these effects can be directly deduced from the chiral and gravitational anomalies. The 
chiral anomaly \cite{anomaly} describes the violation of the classically conserved chiral symmetry 
at the quantum level. It should be noted that in the presence of an external magnetic field the 
chiral anomaly is generated entirely on the lowest Landau level \cite{Ambjorn}. It is essential 
that the operator relation for the chiral anomaly is one-loop exact and cannot get any higher-order 
radiative corrections \cite{Bardeen}. Since the chiral anomalous effects in the free theory are 
generated by the quantum anomalies, it was argued in Refs.~\cite{Zhitnitsky,Newman} 
that the one-loop results for the anomalous transport coefficients should be exact. One should 
keep in mind, however, that in order to get these anomalous transport coefficients one should 
calculate the ground state expectation values of the corresponding operator relations. 
Therefore, {\it a priori} there is no guarantee that interaction corrections should be absent.

The first studies of the interaction effects were done in Refs.~\cite{chiral-shift-1,Ruggieri,chiral-shift-3} 
in the framework of the dense Nambu--Jona-Lasinio (NJL) model in a magnetic field. Using the 
method of Schwinger--Dyson equation for the fermion propagator, it was found that the interaction 
between the fermions in LLL and higher Landau levels promotes the chiral asymmetry 
from the LLL to all Landau levels \cite{chiral-shift-2}. For a magnetic 
field directed along the positive $z$-axis, dynamically generated chiral shift parameter $\Delta$ 
enters the effective action as the $\Delta\bar{\psi}\gamma^3\gamma^5\psi$ term and produces an 
additional dynamical contribution to the axial current. It should be emphasized that since this term 
does not break any symmetry in dense relativistic matter in a magnetic field, the chiral shift is already 
generated in the perturbation theory \cite{chiral-shift-1,chiral-shift-2}. In the NJL model the chiral shift takes  
a constant value independent of the momentum and the Landau level index. In the chiral limit,
it determines the relative shift of the momenta in the dispersion relations of opposite chirality fermions 
$k^3 \to k^3 \pm \Delta$, where the momentum $k^3$ is directed along the magnetic field. 
In other words, it splits the Dirac point into two Weyl nodes separated in momentum space by $2\Delta$. 
It was proposed, therefore, that the same mechanism should take place in Dirac semimetals 
at nonzero charge density: in a magnetic field they transform into Weyl semimetals \cite{engineering}.

Direct quantum-field theoretical calculations performed in dense QED to the leading order in coupling 
and the external magnetic field \cite{radiative-corrections} showed that the axial current in the 
CSE receives nontrivial radiative corrections. It was found that, in the weak-field limit, the radiative 
corrections to the CSE originate from the singularities at the Fermi surface. (The radiative 
corrections to the chiral vorticity conductivity connected with the CVE were calculated in 
Refs.~\cite{Golkar,Ren}.) The role of the interaction effects and radiative corrections in 
various chiral anomalous effects in magnetized relativistic matter were recently discussed 
in Refs.~\cite{Jansen,Zakharov}.

By calculating the electron self-energy in magnetized QED plasma to the leading order in the 
coupling constant and the external magnetic field, it was found in Ref.~\cite{chiral-shift-QED} 
that the chiral asymmetry of the normal ground state of the system is characterized by two distinct 
Dirac structures. While one of them is the chiral shift familiar from the NJL model studies, the 
other Dirac structure is new. It formally looks like that of the chiral chemical potential but is an 
odd function of the longitudinal component of the momentum (i.e., directed along the magnetic field). 
The origin of this new parity-even chiral structure is directly connected with the long-range
character of the QED interaction. The calculations in Ref.~\cite{chiral-shift-QED} were performed 
in the weak magnetic field approximation, using the pseudomomentum representation.
Recently, the same pseudomomentum representation was also tested in the problem of chiral 
symmetry breaking in QCD at zero baryon density \cite{Watson:2013ghq}.

The calculation of the fermion self-energy (as well as the chiral asymmetry functions) 
in the weak-field limit \cite{chiral-shift-QED} revealed an infrared logarithmic singularity. 
This feature may well be an artifact of the expansion in powers of the magnetic field. 
It may also be related to yet another problem. As shown in Ref.~\cite{radiative-corrections}, 
the weak-field result for the axial current density depends on the photon mass which is 
introduced as an infrared regulator. It was argued, however, that such a dependence is
fictitious and is expected to go away after taking into account the nonperturbative 
corrections beyond the weak-field limit.  As a first step in the direction of resolving 
the limitations of the weak-field expansion \cite{chiral-shift-QED}, in this paper we 
study the fermion self-energy and chiral asymmetry in cold magnetized QED plasma 
in the Landau level representation.

This paper is organized as follows. In Sec.~\ref{Sec:Model}, we briefly introduce the model and 
notations. The definition of the fermion self-energy and its Landau-level representation are 
reviewed in Sec.~\ref{Sec:SelfEnergy}. In the same section, we also define the chiral asymmetry 
functions and discuss their ultraviolet properties. The numerical algorithm for calculating the chiral 
asymmetry functions, as well as the main results are presented in Sec.~\ref{Sec:NumericalResults}. 
In Sec.~\ref{Sec:Discussion}, we summarize our findings and give our conclusions. In several 
appendices at the end of the paper, we provide some technical details and derivations used 
in the main text.

\section{Model}
\label{Sec:Model} 

Following closely the notation of Ref.~\cite{chiral-shift-QED}, we start from the following 
Lagrangian density of QED in an external magnetic field:
\begin{equation}
{\cal L}=-\frac{1}{4}F^{\mu\nu}F_{\mu\nu}+\bar{\psi}\left( i\gamma^{\nu}{\cal D}_{\nu}+\mu
\gamma^0-m\right)\psi,
\label{Lagrangian}
\end{equation}
where ${\cal D}_{\nu} =\partial_{\nu}-i e A^{\rm ext}_{\nu}-i e A_{\nu}$ is the covariant derivative,
$\mu$ is the fermion chemical potential, and $m$ is the bare fermion mass. Note that the notation 
is similar to that of Ref.~\cite{Peskin}, but assumes the opposite sign of the electric charge $e$, 
i.e., our $e$ is positive.
Without loss of generality, we assume that the external magnetic field $\mathbf{B}$ points in the $z$ 
direction. The components of the spatial vectors, including those of the vector potential $\mathbf{A}^{\rm ext}$, 
are identified with the {\em contravariant} components. The components of the gradient 
$\bm{\nabla}$ are given by $\partial_k\equiv -\partial^k $. When the explicit form of the 
vector potential is needed, we utilize the Landau gauge, $\mathbf{A}^{\rm ext}= (0, x B,0)$.

In the presence of a constant magnetic field $\mathbf{B}$, part of the translational symmetry in 
the system is broken. This is obvious because, for one-particle states of charged fermions, the 
momentum perpendicular to the magnetic field is not a good quantum number. The absence 
of the translational invariance is reflected in the structure of the fermion propagator 
\cite{Schwinger:1951nm}, i.e., 
\begin{equation}
S(u,u^\prime)=e^{i\Phi(\mathbf{r},\mathbf{r}^\prime)}\bar{S}(u-u^\prime),
\label{propagator-Phi-S}
\end{equation}
where $u=(t,\mathbf{r})$ is a space-time four-vector, $\mathbf{r}=(x,y,z)$,
$\Phi(\mathbf{r},\mathbf{r}^\prime)=-eB(x+x^\prime)(y-y^\prime)/2$ 
is the Schwinger phase, and $\bar{S}(u-u^\prime)$ is the translation invariant part of the 
propagator. A similar form is valid for the inverse fermion propagator as well 
\cite{chiral-shift-QED}, i.e., 
\begin{equation}
S^{-1}(u,u^\prime)=e^{i\Phi(\mathbf{r},\mathbf{r}^\prime)}\bar{S}^{-1}(u-u^\prime).
\label{inverse-propagator-Phi-S}
\end{equation}
It should be emphasized, though, that the translation invariant part $\bar{S}^{-1}(u-u^\prime)$ 
is not the inverse of $\bar{S}(u-u^\prime)$.

\section{Fermion self-energy}
\label{Sec:SelfEnergy}

As proposed in the previous study \cite{chiral-shift-QED}, the chiral asymmetry of the dense QED 
in a magnetic field is captured by the structure of the fermion self-energy. To leading order in 
coupling constant $\alpha=e^2/(4\pi)$, the corresponding expression for the self-energy in
coordinate space reads 
\begin{equation}
\Sigma(u,u^\prime)=-4i\pi\alpha\gamma^\mu S(u,u^\prime) \gamma^\nu D_{\mu\nu}(u-u^\prime).
\label{self-energy}
\end{equation}
By taking into account the structure of the propagator in Eq.~(\ref{propagator-Phi-S}), we find 
that the self-energy has the same Schwinger phase factor, i.e., $\Sigma(u,u^\prime)=e^{i\Phi(\mathbf{r},\mathbf{r}^\prime)}\bar{\Sigma}(u-u^\prime)$. 
After dropping the corresponding phase on both sides of Eq.~(\ref{self-energy}) and performing 
the Fourier transform, we arrive at the following pseudomomentum representation for the 
translation invariant part of the self-energy:
\begin{equation}
\bar{\Sigma}(p)
  =  -4i \pi \alpha \int \frac{dk_0 dk_3 d^2 \mathbf{k}_{\perp}}{(2\pi)^4} \gamma^\mu \bar{S}(k) 
     \gamma^\nu D_{\mu\nu}(k-p) .
\label{self-energy-k-space}
\end{equation}
Here $\bar{S}(k)$ is the Fourier transform of the translation invariant part of the fermion propagator,
and $D_{\mu\nu}(q)$ is the momentum space representation for the photon propagator. 

In the study at hand, we are interested in properties of cold QED matter at nonzero density. 
Moreover, we assume that the fermion number density is large, i.e., the corresponding 
value of the chemical $\mu$ is much larger than other energy scales in the problem.
In particular, we assume that $\mu\gg \sqrt{|eB|}\gg m$, which is a reasonable hierarchy, for
example, in the case of electron plasma in magnetars. One of the most important effects associated 
with the nonzero density is the screening of the one-photon exchange interaction. Even at 
weak coupling, such screening is strong and plays an important role in the dynamics. The 
well-known scheme that captures the corresponding effects is called the hard-dense-loop (HDL) 
approximation \cite{Vija,Manuel}. The explicit form of the HDL photon propagator in Euclidean 
space is given by 
\begin{equation} 
 D_{\mu\nu}(q) \simeq i\left(\frac{|\bm{q}| }{|\bm{q}|^3+\frac{\pi}{4}m_D^2|q_{4}|}O^{\rm (mag)}_{\mu\nu}
+ \frac{O^{\rm (el)}_{\mu\nu}}{q_{4}^2+|\bm{q}|^2+m_D^2}\right),
\label{D-HDL}
\end{equation}
where $q_4\equiv iq_0$ and $m_D^2=2\alpha \mu^2/\pi$ is the Debye screening mass.
In the Coulomb gauge assumed here, the Lorentz space projectors onto the electric and 
magnetic modes are defined as follows:
\begin{eqnarray} 
O^{\rm (mag)}_{\mu\nu} &=&  g_{\mu\nu}-u_{\mu} u_{\nu}
+\frac{\bm{q}_{\mu}\bm{q}_{\nu}}{|\bm{q}|^{2}},\\
O^{\rm (el)}_{\mu\nu} &=&  u_{\mu} u_{\nu},
\end{eqnarray}
where $u_{\mu}=(1,0,0,0)$. 

The explicit form of the translation invariant part of the free fermion propagator is given by \cite{Chodos:1990vv}
\begin{equation}
\bar{S}(k) = 2 i  e^{-k_{\perp}^2 l^2 }  \sum_{n=0}^{\infty} \frac{(-1)^n 
D_n(k)}{[k_0+\mu+i\epsilon\,\mbox{sgn}(k_0)]^2-2n|eB| -(k^3)^2-m^2},
\label{prop-momentum}
\end{equation}
where $l\equiv 1/\sqrt{|eB|}$ is the magnetic length and 
\begin{eqnarray}
D_n(k) &=& \left[\gamma^0 (k_0+\mu) -\gamma^3 k^3 +m\right]
\left[ L_{n}\left(2k_{\perp}^2 l^2\right) {\cal P}_{-} 
-  L_{n-1}\left(2k_{\perp}^2 l^2\right){\cal P}_{+}\right] 
 +2(\bm{\gamma}_{\perp}\cdot\mathbf{k}_\perp) L^{1}_{n-1}\left(2k_{\perp}^2 l^2\right).
\label{prop-D-n-momentum}
\end{eqnarray}
Here ${\cal P}_{\pm}=\left[1\pm i\,\mbox{sgn}(eB)\gamma^1\gamma^2\right]/2$ are spin projectors 
and $L^{(\alpha)}_n(x)$ are generalized Laguerre polynomials \cite{Gradstein_Ryzhik}. By definition,
$L_{-1}(x)=0$.

The structure of the one-loop self-energy (\ref{self-energy-k-space}) was discussed in detail in 
Ref.~\cite{chiral-shift-QED} by utilizing the Landau-level representation, recently developed 
in Ref.~\cite{Gorbar:2011kc}. Just like the fermion propagator in Eqs.~(\ref{prop-momentum}) 
and (\ref{prop-D-n-momentum}), the self-energy can be expanded over the Landau levels.
The corresponding general form reads 
\begin{eqnarray}
\bar{\Sigma}(p)&=& 2  e^{-p_{\perp}^2 l^2} 
\sum_{n=0}^{\infty}(-1)^n\Big\{
\left(-\gamma^0 \delta\mu_{n} 
+p^3\gamma^3  \delta v_{3,n}
- i\gamma^1\gamma^2\tilde{\mu}_{n}
-\gamma^3\gamma^5\Delta_{n}
-\gamma^0\gamma^5\mu_{5,n}
+{\cal M}_{n} \right) \nonumber\\
&&\times \left[ L_{n}(2p_{\perp}^2 l^2){\cal P}_{-} - L_{n-1}(2p_{\perp}^2 l^2){\cal P}_{+}\right]
-2 (\bm{\gamma}_\perp\cdot \bm{p}_{\perp}){\delta v_{\perp,n}} L^{1}_{n-1}(2p_{\perp}^2 l^2)
\Big\}.
\label{self-energy-LL}
\end{eqnarray} 
In this representation, the physical meaning of the coefficient functions $\delta\mu_n$, $\delta v_{3,n}$, 
etc., is obvious from their Dirac structure \cite{chiral-shift-QED}. [Note that all these functions depend 
on the energy $p_0$ and the longitudinal momentum $p_3$.] In the remainder of this study, however, 
we will concentrate only on the two most important structures, $\Delta_{n}(p_3)$ and $\mu_{5,n}(p_3)$
at $p_0=0$, which describe the chiral asymmetry of dense QED matter. 
General expressions for both of these were derived in Ref.~\cite{chiral-shift-QED} 
by projecting the self-energy in Eq.~(\ref{self-energy-k-space}) onto individual Landau levels, i.e.,
\begin{eqnarray}
\Delta_n(p_3) & = & \frac{(-1)^n}{8} \frac{l^2}{\pi} \mathrm{sign}(eB) \int d^2\mathbf{p}_\perp e^{-p^2_{\perp}l^2} 
                    \left[  L_n(2p_{\perp}^2l^2) + L_{n-1}(2p_{\perp}^2l^2) \right] 
                    \mathrm{Tr} \left[ \gamma^0 \bar{\Sigma}( p) \right] \nonumber \\
          &  -  & \frac{(-1)^n}{8} \frac{l^2}{\pi} \int d^2\mathbf{p}_\perp e^{-p^2_{\perp}l^2} 
                    \left[  L_n(2p_{\perp}^2l^2) - L_{n-1}(2p_{\perp}^2l^2) \right] 
                    \mathrm{Tr} \left[ \gamma^3 \gamma^5 \bar{\Sigma}( p) \right],  
\label{Delta-n}
\end{eqnarray}
and 
\begin{eqnarray}
\mu_{5,n}(p_3) & = & \frac{(-1)^n}{8} \frac{l^2}{\pi} \mathrm{sign}(eB) \int d^2\mathbf{p}_\perp e^{-p^2_{\perp}l^2} 
                    \left[  L_n(2p_{\perp}^2l^2) + L_{n-1}(2p_{\perp}^2l^2) \right] 
                    \mathrm{Tr} \left[ \gamma^3 \bar{\Sigma}( p) \right] \nonumber \\
          &  +  & \frac{(-1)^n}{8} \frac{l^2}{\pi} \int d^2\mathbf{p}_\perp e^{-p^2_{\perp}l^2} 
                    \left[  L_n(2p_{\perp}^2l^2) - L_{n-1}(2p_{\perp}^2l^2) \right] 
                    \mathrm{Tr} \left[ \gamma^0 \gamma^5 \bar{\Sigma}( p) \right],  
\label{mu-5-n}
\end{eqnarray}
respectively.

At large chemical potential considered here, magnetic catalysis \cite{Gusynin:1995gt} 
plays no role and the dynamical contribution to ${\cal M}_{n}$ is negligible even compared 
to the electron mass $m_e$. It is completely justifiable, therefore, to replace ${\cal M}_{n}$ 
with $m$ in our calculations below. Moreover, the dynamical contribution to the electron 
mass due to the magnetic catalysis in QED is exponentially small even in the case of zero 
chemical potential (vacuum) \cite{Gusynin:1995gt,Leung:1996qy,Gusynin:1998zq}. This is 
the consequence of the smallness of the fine structure constant. While the same is not 
true in the QCD vacuum, the magnetic catalysis still would not play any big role in dense 
quark matter at large chemical potential.

Before proceeding to the numerical analysis of the chiral asymmetry functions $\Delta_{n}(p_3)$ 
and $\mu_{5,n}(p_3)$, let us discuss the implications of the well-known ultraviolet
divergency in the fermion self-energy function in QED \cite{Peskin}. In the Pauli-Villars 
regularization scheme, the self-energy contains the following logarithmically divergent 
contribution \cite{Peskin}:
\begin{equation}
\Sigma^{\rm (div)} (p) = \frac{\alpha}{4\pi} \left[-\gamma^\nu (p_\nu+\mu\delta_{\nu}^{0}) +4m \right]
\ln\frac{\Lambda^2}{m^2} .
\end{equation}
Note that the only effect of a nonzero chemical potential here is to shift $p_0\to p_0+\mu$ 
in the vacuum expression \cite{radiative-corrections,Freedman:1976xs}. Of course, the 
above divergency cannot be affected by the magnetic field. When projected onto 
Landau levels as prescribed by Eqs.~(\ref{Delta-n}) and (\ref{mu-5-n}), this result 
leads to the following contributions to the chiral asymmetry functions:
\begin{eqnarray}
\Delta^{\rm (div)}_n(p_3) & = & - \delta_{n}^{0}\frac{\alpha (p_0+\mu)  }{8\pi} \mathrm{sign}(eB) \ln\frac{\Lambda^2}{m^2} ,\\
\mu^{\rm (div)}_{5,n}(p_3) & = &  \delta_{n}^{0} \frac{\alpha p_3 }{8\pi} \mathrm{sign}(eB)  \ln\frac{\Lambda^2}{m^2} .
\end{eqnarray}
As we see, the corresponding functions are free of the ultraviolet divergences in all, but the lowest Landau level 
($n=0$). As explained in detail in Ref.~\cite{chiral-shift-QED}, the LLL ($n=0$) is very special also because of 
its spin-polarized nature. As a consequence, the LLL chiral shift is indistinguishable from the correction 
to the chemical potential, and the LLL axial chemical potential is indistinguishable from the correction 
to the longitudinal velocity. It was concluded, therefore, that the novel type of the chiral asymmetry is determined 
exclusively by the dynamical functions $\Delta_n$ and $\mu_{5,n}$ with $n \ge 1$. These functions 
are of prime interest for us in the present paper. In the next section, we take into account that all 
these functions are free from the ultraviolet divergences and study them numerically.

\section{Numerical results for chiral asymmetry}
\label{Sec:NumericalResults}

In this section we study the chiral asymmetry functions $\Delta_{n}(p_3)$ and $\mu_{5,n}(p_3)$ using 
numerical methods. To start with, we rewrite the corresponding expressions in a dimensionless
form. We will measure all quantities with the dimension of energy/mass in terms of the chemical potential. 
For example, in the case of momenta, we will define the corresponding dimensionless quantities as follows: 
$x \equiv p_{\perp}/\mu$, $y \equiv  k_{\perp}/\mu$,  $ x_{3} \equiv p_{3}/\mu$, and $y_{3} \equiv k_{3}/\mu$. 
Similarly, the dimensionless functions will be defined as follows:
$\bar{\Delta}_n \equiv  \Delta_n/\mu$ and $\bar{\mu}_{5,n} \equiv  \mu_{5,n}/\mu$. 
The corresponding dimensionless forms of these chiral asymmetry functions are presented 
in Appendix~\ref{App:Chiral-asymmetry-functions}.

In order to analyze numerically the two chiral asymmetry functions in dense QED, we need 
to fix several model parameters (i.e., the strength of magnetic field, the value of the chemical potential,
and the fermion mass). In principle, when using the dimensionless description, the value of the 
chemical potential $\mu$ may be left unspecified. Keeping in mind, however, that the value of 
the fermion (electron) mass has to be measured in units of $\mu$, we will assume that the 
default choices of the chemical potential and the magnetic field are $\mu=420~\mbox{MeV}$ 
and $B=10^{18}~\mbox{G}$. Then, the two dimensionless model parameters used in the 
calculations will be
\begin{eqnarray}
a & = & \frac{m}{\mu} \approx 1.22\times10^{-3} \frac{m}{m_e} \frac{420~\mbox{MeV}}{\mu}, 
\label{def-a}\\
b & = & \frac{ |eB| }{\mu^2} \approx \frac{1}{30} \left( \frac{B}{10^{18}~\mbox{G}} \right) \left( \frac{420~\mbox{MeV}}{\mu} \right)^2.
\label{def-b}
\end{eqnarray}
Note that, in agreement with the assumption made earlier, the chosen value of the magnetic field 
strength is rather small compared to the chemical potential scale $\mu^2$. By taking into account 
the definition of the Debye mass and the QED fine structure constant, it is also convenient to  
introduce the following short-hand notation for the dimensionless Debye mass:
\begin{equation}
d = \frac{m_D}{\mu} \equiv \sqrt{ \frac{2\alpha}{\pi} } \approx 6.816\times10^{-2}.
\label{def-d}
\end{equation}
In the final expressions for the chiral asymmetry functions, there will be a need to sum over  
an infinite number of Landau levels. In the numerical calculations, however, the sums will 
be truncated at $n_{\rm max}=200$. 

As is clear from the explicit expressions for the functions $\bar{\Delta}_n$ and $\bar{\mu}_{5,n}$ 
in Appendix~\ref{App:Chiral-asymmetry-functions} [see Eqs.~(\ref{Delta-n-dimless}) and 
(\ref{mu-5-n-dimless}), respectively], the numerical calculation for each of them reduces 
to performing four integrations: three integrations over the dimensionless momenta 
$x$, $y$ and $y_3$, and one over the angular coordinate $\phi$. Taking into account 
that $\bar{\Delta}_n$ and $\bar{\mu}_{5,n}$ also have an additional functional dependence 
on the longitudinal momentum $x_{3} \equiv p_{3}/\mu$, the corresponding task becomes 
rather expensive numerically. 

Before proceeding to the actual results, let us briefly describe the algorithm that we use
in the calculations. The four-dimensional integrals that define the chiral asymmetry functions
have the following schematic form:
\begin{equation}
I=\int_{0}^{\infty} dx\int_{0}^{\infty} dy\int_{-\infty}^{\infty} dy_3 \int_{0}^{2\pi} \frac{d\phi}{2\pi}
f(x,y,y_3,\phi).
\end{equation}
In order to calculate such an integral, we will make use of the importance sampling Monte Carlo method 
\cite{Weinzierl:2000wd}. In such a framework, the result of the integration is approximated by a 
weighted sum of contributions calculated at a large number of random points in the phase space, 
i.e.,
\begin{equation}
I_{N}=\frac{1}{N}\sum_{i=1}^{N} \frac{f(x^{i},y^{i},y_3^{i},2\pi u^{i})}{P_1(x^{i}) P_2(y^{i}) P_3(y_3^{i})}.
\end{equation}
Having limited information about the angular dependence of the integrand function, we use the 
simplest uniform distribution of the random variable $u = \phi/(2\pi)$ on the interval from 
$0$ to $1$.
The other three random number variables are distributed with the probability density functions 
$P_1(x)$, $P_2(y)$, and $P_1(y_3)$, respectively. The specific choice of these functions will 
be explained momentarily. First, however, let us note that the statistical error of the Monte 
Carlo integration is given by the following estimator \cite{Weinzierl:2000wd}:
\begin{equation}
\epsilon = \frac{1}{\sqrt{N}}\sqrt{\frac{1}{N}\sum_{i=1}^{N}
\left(\frac{f(x^{i},y^{i},y_3^{i},2\pi u^{i})}{P_1(x^{i}) P_2(y^{i}) P_3(y_3^{i})}\right)^2-I_{N}^2}.
\label{estimator}
\end{equation}
With increasing $N$, the Monte Carlo estimate $I_{N}$ may converge rather slowly to the true 
value $I$. This is where importance sampling can improve the situation. The key observation is 
that, for a fixed number of sampling points $N$, the result for the above statistical error 
depends on the random number distributions used. The error becomes smaller when the 
corresponding probability densities approximate closer the integrand function itself.
The same condition determines when the fastest convergence of the Monte Carlo 
method is achieved.

While testing our numerical algorithm, we tried performing calculations with several different types 
of the functional forms for the random number distributions (e.g., Gaussian and gamma distributions) 
and examined a number of different choices of their parameters. In such tests, the least value of 
the estimator (\ref{estimator}) was used as an indicator of the integration effectiveness. This allowed  
us to make an optimal choice of the random number distributions.  

In the case of the perpendicular momenta variables $x$ and $y$, we ended up using the following 
gamma distribution:
\begin{equation}
P(x)=\frac{\beta^\alpha}{\Gamma(\alpha)} x^{\alpha-1}e^{-\beta x},
\end{equation}
where the shape and scale parameters are $\alpha=1$ and $\beta=1.5$, respectively. 
Such a distribution appears indeed appropriate in the case of the integrand function 
that decreases exponentially with the perpendicular momenta. In order to generate 
gamma-distributed random numbers, we used the {\em FORTRAN} 
code written by Richard Chandler \cite{gamma-distribution}. 

As a quick examination of the explicit expressions for $\bar{\Delta}_n$ and $\bar{\mu}_{5,n}$ 
in Appendix~\ref{App:Chiral-asymmetry-functions} reveals, the dependence of their 
integrands on the longitudinal momentum $y_3$ is quite different from the dependence
on $x$ and $y$. In particular, they have a power-law instead of exponential behavior at 
large $y_3$. Because of this, neither Gaussian nor gamma distributions were able to 
provide a quick convergence of the Monte Carlo integration. Instead, we used the 
Cauchy distribution with the following power-law probability density function for 
generating the longitudinal momentum variable $y_3$:
\begin{equation}
P_3(y_3)=\frac{1}{\pi} \frac{1}{y_3^2+1}.
\end{equation}
The random numbers with such a probability density are generated using  
the quantile function $y_3=\tan\left[\frac{\pi}{2}(2p -1)\right]$, where $p$ is
a random number with the uniform distribution on the interval between $0$ 
and $1$. 

\begin{figure}[ht]
\centering
\includegraphics[width=0.47\textwidth]{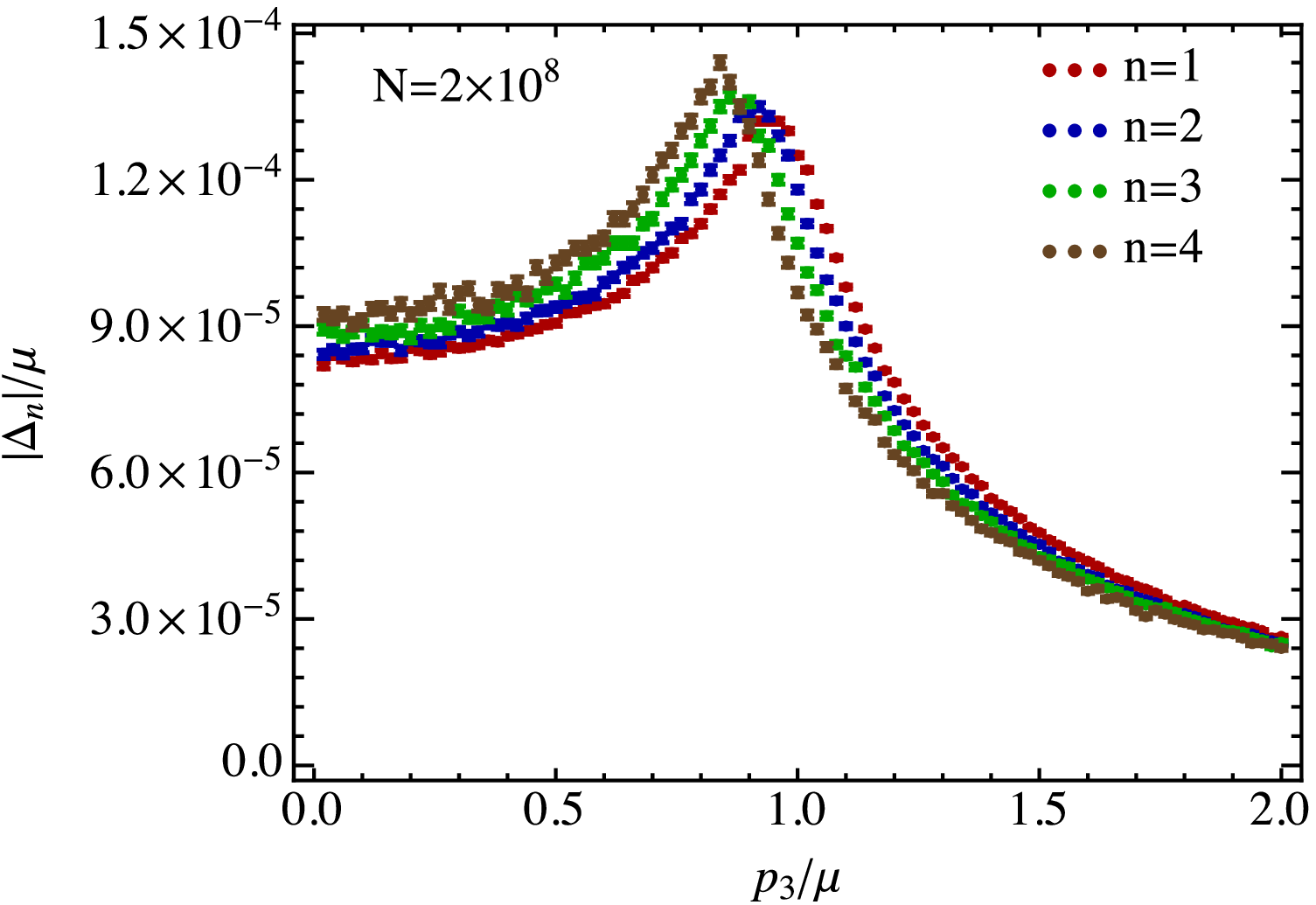}\hspace{0.03\textwidth}
\includegraphics[width=0.47\textwidth]{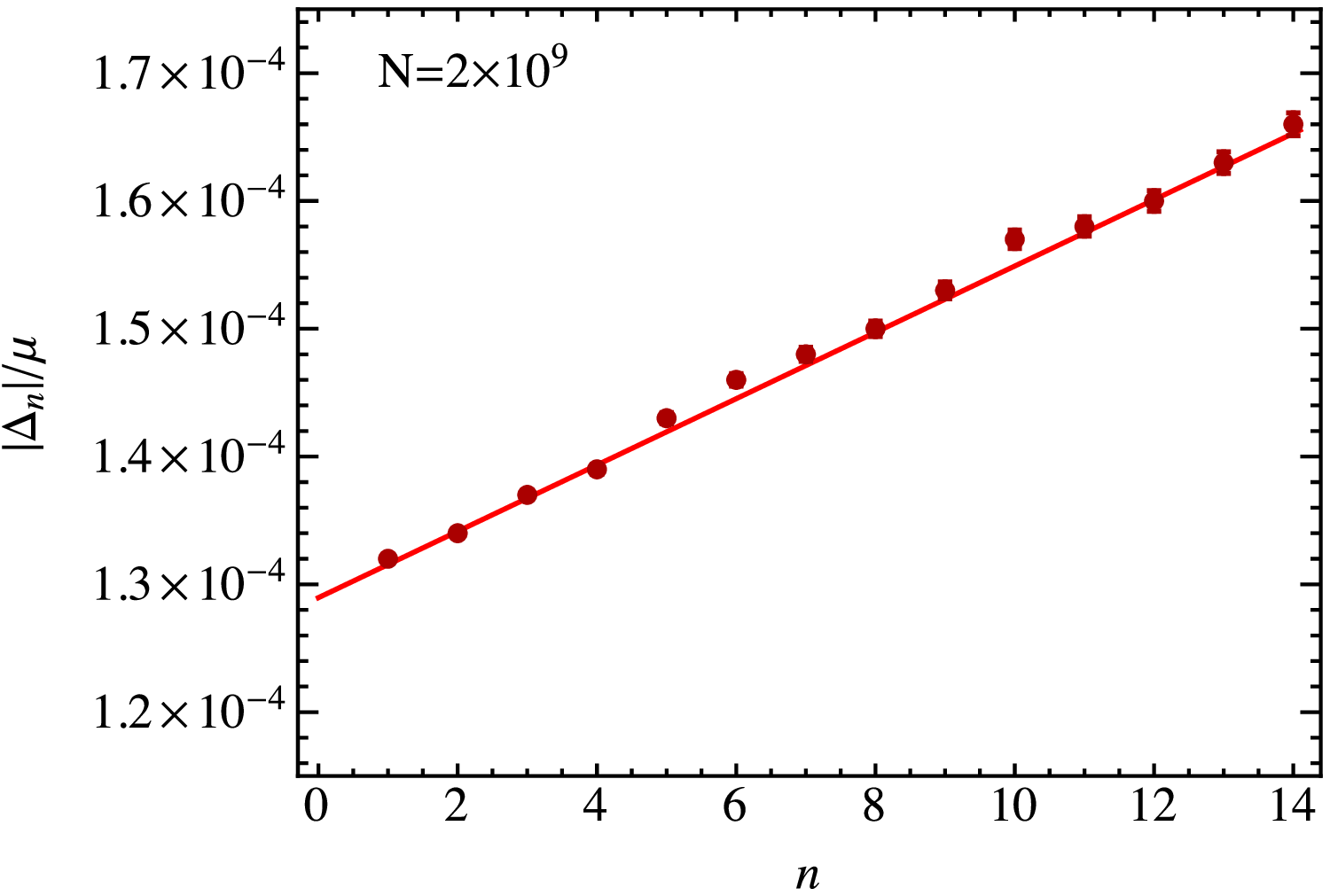}
\caption{(color online) 
Left panel: the chiral shift $\Delta_n$ as a function of the longitudinal 
momentum $p_3$ for $n=1$ (red), $n=2$ (blue), $n=3$ (green), and $n=4$ (brown) Landau 
levels. Right panel: the values of the chiral shift $\Delta_n$ at the Fermi surface.}
\label{fig_Delta-n}
\end{figure}

Our numerical results for the chiral shift are summarized in Fig.~\ref{fig_Delta-n}.
In the left panel, we show the dependence of the chiral shift $\Delta_{n}$ on the 
longitudinal momentum $y_3=p_3/\mu$ for several low-lying Landau levels. 
Since obtaining the complete functional dependence on $p_3$ is rather 
expensive numerically, we used only a moderately large number of sampling
points, $N=2\times 10^8$ and calculated the results only for the first four lowest 
lying Landau levels. The common feature of the corresponding functions 
is the appearance of a maximum at an approximate location of the Fermi surface. 
In the free (weakly interacting) theory, this is determined by the following value of
the longitudinal momentum: $p_3/\mu = \sqrt {1 - 2 nb - a^2}$, where $a$ and $b$ 
are defined in Eqs.~(\ref{def-a}) and (\ref{def-b}). In agreement with 
this expression, the location of the maximum of the chiral shift function in the $n$th 
Landau level $\Delta_{n}(p_3)$ decreases with increasing $n$. At large values of 
the momentum $p_3$, the chiral shift function decreases and gradually approaches 
zero as expected. 

From the viewpoint of the low-energy physics, it is most important to know the chiral 
shift at the Fermi surface. The corresponding results are presented in the right 
panel of Fig.~\ref{fig_Delta-n}. By assumption, the location of the Fermi surface
is determined by the perturbative expression, $p_3/\mu = \sqrt {1 - 2 nb - a^2}$. 
In this calculation, we used a larger number of sampling points, $N=2\times 10^9$. 
As we see, the Fermi surface values of $\Delta_{n}$ grow with the Landau-level 
index $n$. (The corresponding numerical values are also given in the first 
column of Table~\ref{Table-Fermi-surface}.) This growth is somewhat surprising, 
but is in agreement with the general behavior of functions $\Delta_{n}(p_3)$ 
shown in the left panel of Fig.~\ref{fig_Delta-n}. The corresponding dependence 
on the Landau-level index can be fitted quite well by a linear function. 

It is easy to check that the numerical results for the chiral shift in Fig.~\ref{fig_Delta-n} 
are of the same order of magnitude as $\alpha |eB|/\mu$. Taking into account that 
$\Delta_{n}$ is one of the structures in the fermion self-energy, induced by a nonzero 
magnetic field, it is indeed quite natural that the corresponding function is proportional 
to the coupling constant and the magnetic field strength. As for the chemical potential 
in the denominator, it is the only other relevant energy scale in the problem that can 
be used to render the result for $\Delta_{n}$ with the correct energy units. 
(Formally, the fermion mass is yet another energy scale, but it is unlikely 
to play a prominent role at the Fermi surface in the high density and strong-field 
limit.) The linear fit for the chiral shift at the Fermi surface is shown by the solid 
line in the right panel of Fig.~\ref{fig_Delta-n}. The corresponding function can 
be presented in the following form:
\begin{equation}
\Delta_{n} \simeq -\frac{\alpha |eB|}{\mu}\left(0.53+0.32 \frac{|eB|n}{\mu^2}\right),
\label{Delta-n-fit}
\end{equation}
where we took into account that the numerical results in Fig.~\ref{fig_Delta-n} were 
obtained for the magnetic field $|eB|=\mu^2/30$ and $\alpha =1/137$. The result in 
Eq.~(\ref{Delta-n-fit}) should be contrasted with a very different parametric dependence 
obtained in the weak-field limit in Ref.~\cite{chiral-shift-QED}, i.e., $\Delta_{n} \propto 
\alpha |eB|\mu/m^2$, which is a factor of $(\mu/m)^2$ larger. Such a large factor is 
quite natural in the weak-field limit, where it is an artifact of the expansion in powers of 
$|eB|/m^2$. In contrast, one does not expect anything like that in the case of a strong 
magnetic field.

\begin{figure}[ht]
\centering
\includegraphics[angle=0,width=0.47\textwidth]{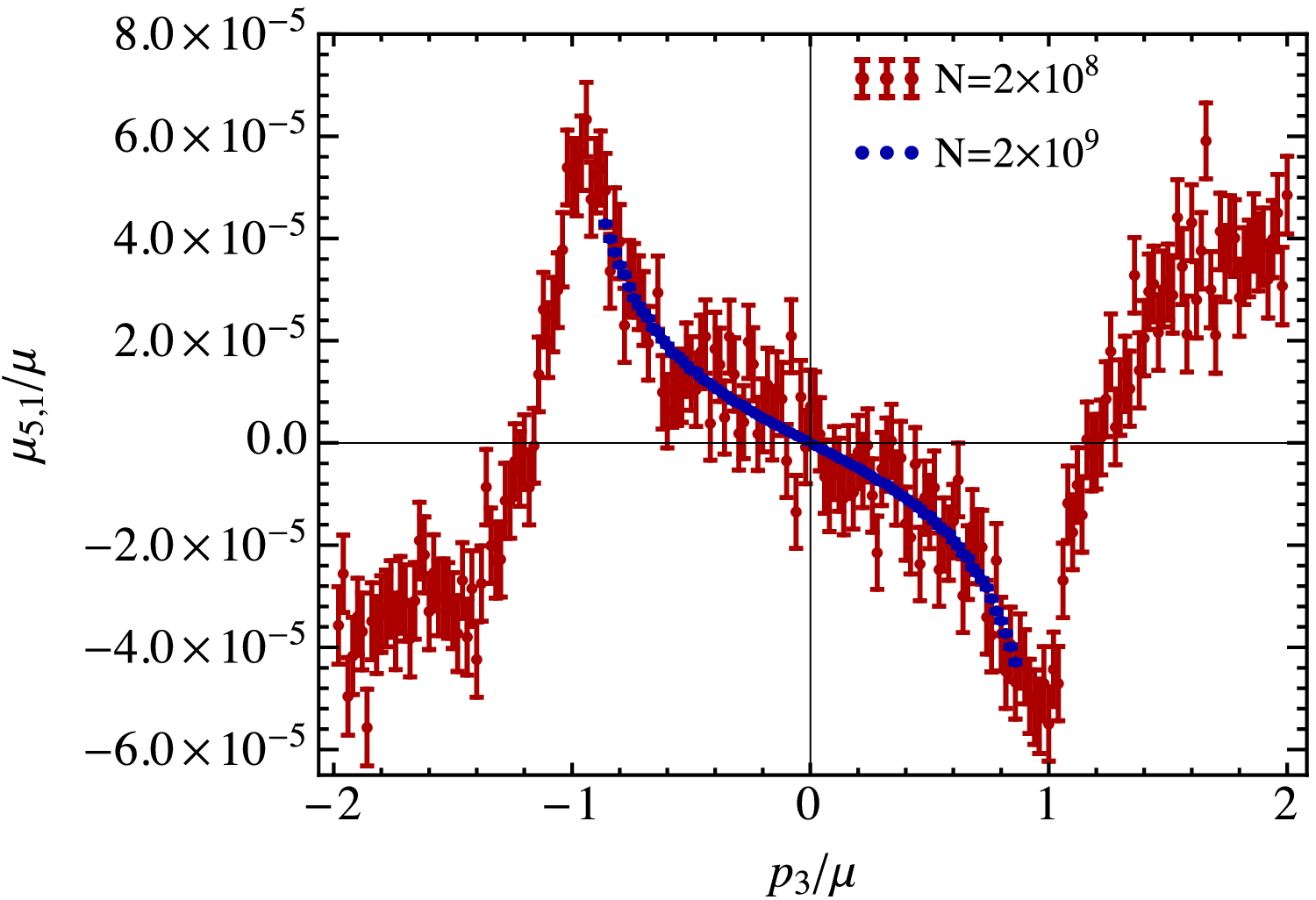}\hspace{0.03\textwidth}
\includegraphics[width=0.47\textwidth]{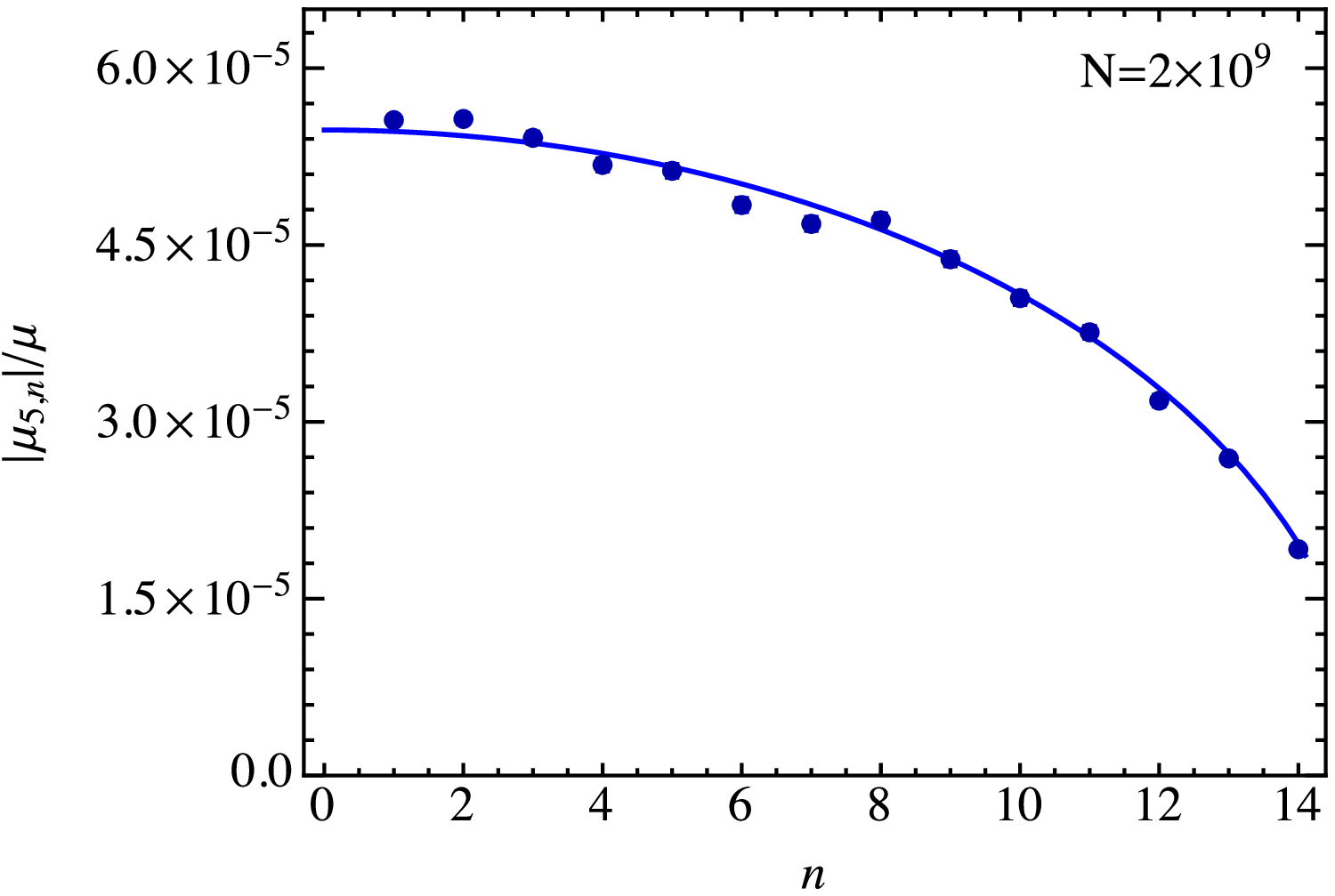}
\caption{(color online) 
Left panel: the chiral chemical potential $\mu_{5,n}$ as a function of the longitudinal 
momentum $p_3$ for $n=1$ Landau level. Right panel: the values of the chiral 
chemical potential $\mu_{5,n}$ at the Fermi surface.}
\label{fig_mu-5-n}
\end{figure}

The numerical results for the chiral chemical potential $\mu_{5,n}$ are summarized in 
Fig.~\ref{fig_mu-5-n}. In the left panel, we present the chiral chemical potential in the 
$n=1$ Landau level as a function of the longitudinal momentum $p_3$. (The results 
for larger $n$ are expected to have the same qualitative dependence on $p_3$.) 
The red and blue points represent the results for two different numbers of sampling 
points, $N=2\times 10^8$ and $N=2\times 10^{9}$, respectively. The numerical 
results confirm that $\mu_{5,n}$ is an odd function of $p_3$ and, as such, it does 
not break parity. The dependence of $\mu_{5,n}$ on $p_3$ also reveals a pair of 
sharp peaks on the Fermi surface at $p_3/\mu \simeq \pm\sqrt {1 - 2 nb - a^2}$. 
In the context of the low-energy physics, it is these values of $\mu_{5,n}$ on the 
Fermi surface that are of main importance.

The numerical results for the chiral chemical potential at the Fermi surface are shown 
in the right panel of Fig.~\ref{fig_mu-5-n}. In the corresponding calculation, we again 
assumed that the location of the Fermi surface is determined by the perturbative 
expression, $p_3/\mu = \pm \sqrt {1 - 2 nb - a^2}$, and used the Monte Carlo integration
algorithm with $N=2\times 10^9$ sampling points. We find that the values of $\mu_{5,n}$ 
decrease with the Landau-level index $n$. (The corresponding numerical values are 
given in the second column of Table~\ref{Table-Fermi-surface}.) The order of magnitude 
of the obtained results is similar to those for the chiral shift function. Following the same
arguments, therefore, we can assume that $\mu_{5,n}$ is also proportional to the 
coupling constant and the magnetic field strengths, i.e., $\mu_{5,n} \propto \alpha |eB|/\mu$.
(Let us emphasize again that this dependence is quite different from the weak-field 
limit in Ref.~\cite{chiral-shift-QED}.) In order to fit the numerical results, we could 
try to use a polynomial function of $n$. However, by following a trial and error approach 
instead, we found that the following simple function approximates our numerical results 
really well: 
\begin{equation}
\mu_{5,n} \simeq - 0.225 \frac{\alpha |eB|}{\mu}\sqrt{1-\left(\frac{2n|eB|}{\mu^2}\right)^2},
\label{mu5-n-fit}
\end{equation}
where we took into account that $|eB|=\mu^2/30$ and $\alpha =1/137$. The corresponding 
function is shown by the solid line in the right panel of Fig.~\ref{fig_mu-5-n}. 

\begin{figure}[ht]
\centering
\includegraphics[width=0.47\textwidth]{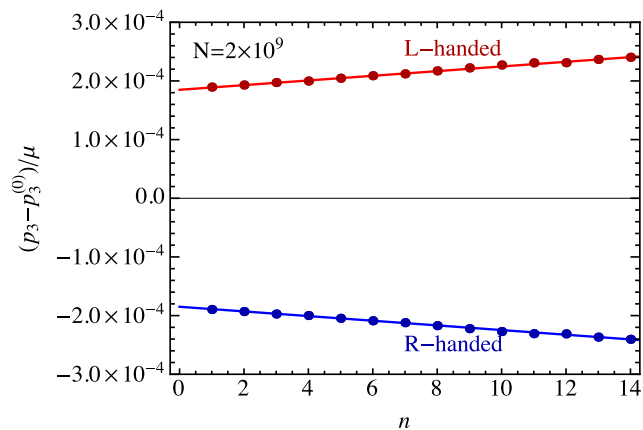}
\caption{(color online) Asymmetry of the Fermi surface for predominantly left-handed (red)
and right-handed (blue) particles as a function of the Landau-level index $n$.}
\label{fig_pFermi}
\end{figure}

By making use of the analytical expression for the fermion propagator with the chiral 
asymmetry in Appendix~\ref{App:Propagator}, as well as the above numerical results 
for the chiral shift and the chiral chemical potential, we can straightforwardly determine 
the interaction-induced deviations of the Fermi momenta $(p_3-p_3^{(0)})/\mu$ for 
the predominantly left-handed and right-handed fermions in the considered 
ultrarelativistic limit $\mu\gg m$. Here $p_3^{(0)}$ is the value of the Fermi 
momentum in the absence of the chiral asymmetry (i.e., $\Delta_{n}=0$ and 
$\mu_{5,n}=0$). Such 
deviations can be viewed as the actual measure of the chiral asymmetry at the 
Fermi surface. The numerical results for $(p_3-p_3^{(0)})/\mu$ in each occupied 
Landau level are shown in Fig.~\ref{fig_pFermi}. (The corresponding numerical 
values are also presented in the last column of Table~\ref{Table-Fermi-surface}.) 
This is a generalization of the analogous results in the weak-field limit, obtained in 
Ref.~\cite{chiral-shift-QED}. 

We find that the results for $(p_3-p_3^{(0)})/\mu$ in Fig.~\ref{fig_pFermi} are well approximated by 
linear functions of $n$. When written in the same form as the chiral shift and the chiral 
chemical potential functions, the corresponding linear fits take the following form:
\begin{equation}
p_3-p_3^{(0)} \simeq \pm \frac{\alpha |eB|}{\mu}\left(0.76+0.49 \frac{|eB|n}{\mu^2}\right).
\label{p3-p30}
\end{equation}
As is easy to check, the values of these Fermi momenta shifts are of the order of
$10$--$100~\mbox{keV}$ and, thus, are not very large in the context of compact stars, even 
though we already assumed an extremely strong value of the magnetic field, 
$B=10^{18}~\mbox{G}$. One should keep in mind, however, that here we used the 
model of a dense QED plasma, whose coupling constant $\alpha$ is extremely small. 
This conclusion could change drastically in the case of dense quark matter, where the 
relevant coupling constant $\alpha_{s}$ is about two orders of magnitude stronger. Indeed, 
by taking into account that the estimate for the Fermi momenta shift in Eq.~(\ref{p3-p30}) 
is proportional to the coupling, we conclude that the chiral asymmetry should be of the 
order of $1$--$10~\mbox{MeV}$ in dense quark matter. Such a large asymmetry may in turn 
produce a substantial neutrino emission asymmetry with observable consequences for 
protoneutron stars \cite{chiral-shift-2}.

\section{Discussion}
\label{Sec:Discussion}

In this paper, we studied the chiral asymmetry induced by a strong external magnetic field in 
cold dense QED matter. This study extends the general predictions of Ref.~\cite{chiral-shift-QED}
regarding the structure of the chiral asymmetry at Fermi surface. Unlike the weak-field 
analysis of Ref.~\cite{chiral-shift-QED}, however, the present paper addressed the 
problem in the general framework that relies on the Landau-level representation. 
Additionally, the screening effects of dense plasma are taken into account in this study.
This is done by utilizing the conventional hard-dense-loop approximation, which is 
justified in the regime considered. 

Among the main results are the numerical functions for the chiral shift and the 
chiral chemical potential. The dependence of both functions on the longitudinal 
momentum reveals local peaks at the approximate position of the Fermi surface.
This feature is in qualitative agreement with the perturbative weak-field results in 
Ref.~\cite{chiral-shift-QED}, where such functions had logarithmic singularities. 

The values of the chiral shift $\Delta_{n}$ and the chiral chemical potential $\mu_{5,n}$ 
at the Fermi surface appear to be of order $\alpha |eB|/\mu$. This differs from the
corresponding weak-field prediction $\alpha |eB|\mu/m^2$ by a rather large factor 
$(\mu/m)^2$. Such a difference is not surprising and, in fact, should have been 
expected in the ultrarelativistic limit when $|eB|/m^2$ is not a good expansion 
parameter. While the dependence of $\Delta_{n}$ on the Landau-level index $n$ 
shows a weak growth, $\mu_{5,n}$ decreases with $n$. By fitting the numerical 
results, we proposed simple model functions which describe the results quite well. 

By making use of our numerical results for $\Delta_{n}$ and $\mu_{5,n}$, we also
obtained the interaction induced deviations of the Fermi momenta $(p_3-p_3^{(0)})/\mu$ 
for the predominantly left-handed and right-handed fermions.
These provide the formal measure of the chiral asymmetry at the Fermi surface. 
The corresponding values appear to be rather small in the case of dense QED 
matter even at extremely large densities and extremely strong magnetic fields.
We suggest, however, that the asymmetry can be substantial in the case of quark
matter.

\acknowledgments
The authors thank ASU Advanced Computing Center for providing computing resources.
The work of E.V.G. was supported partially by SFFR of Ukraine, Grant No.~F53.2/028. 
and by the SCOPES under Grant No.~IZ7370\_152581 of the Swiss NSF. 
The work of V.A.M. was supported by the Natural Sciences and Engineering Research Council of Canada. 
The work of I.A.S. was supported in part by the U.S. National Science Foundation under Grant No.~PHY-1404232.

\begin{table}
\begin{ruledtabular}
\caption{Data for the chiral asymmetry functions $\Delta_n$, $\mu_{5,n}$, and 
$(p_3-p_3^{(0)})/\mu$ at the Fermi surface.} 
\label{Table-Fermi-surface}
\begin{tabular}{cccc}
$n$ & $\Delta_n/\mu$ & $\mu_{5,n}/\mu$ & $(p_3-p_3^{(0)})/\mu$  \\
\hline 
  1  &  $-1.32\times 10^{-4} \pm 1.84\times 10^{-7} $  &  $-5.56\times 10^{-5} \pm 3.30\times 10^{-7}$  &  $\pm 1.90\times 10^{-4}$ \\
  2  &  $-1.34\times 10^{-4} \pm 2.56\times 10^{-7} $  &  $-5.57\times 10^{-5} \pm 4.12\times 10^{-7}$  &  $\pm 1.93\times 10^{-4}$    \\
  3  &  $-1.37\times 10^{-4} \pm 3.21\times 10^{-7}$  &  $-5.41\times 10^{-5} \pm 4.64\times 10^{-7} $  &  $\pm 1.97\times 10^{-4}$    \\
  4  &  $-1.39\times 10^{-4} \pm 3.76\times 10^{-7} $  &  $-5.18\times 10^{-5} \pm 5.01\times 10^{-7}$  &  $\pm 2.00\times 10^{-4}$    \\
  5  &  $-1.43\times 10^{-4} \pm 4.34\times 10^{-7}$  &  $-5.13\times 10^{-5} \pm 5.27\times 10^{-7}$  &  $\pm 2.05\times 10^{-4}$    \\
  6  &  $-1.46\times 10^{-4} \pm 4.90\times 10^{-7}$  &  $-4.84\times 10^{-5} \pm 5.43\times 10^{-7}$  &  $\pm 2.09\times 10^{-4}$    \\
  7  &  $-1.48\times 10^{-4} \pm 5.46\times 10^{-7}$  &  $-4.68\times 10^{-5} \pm 5.52\times 10^{-7}$  &  $\pm 2.12\times 10^{-4}$    \\
  8  &  $-1.50\times 10^{-4} \pm 5.97\times 10^{-7}$  &  $-4.71\times 10^{-5} \pm 5.52\times 10^{-7}$  &  $\pm 2.17\times 10^{-4}$    \\
  9  &  $-1.53\times 10^{-4} \pm 6.54\times 10^{-7}$  &  $-4.38\times 10^{-5} \pm 5.42\times 10^{-7}$  &  $\pm 2.22\times 10^{-4}$    \\
 10  &  $-1.57\times 10^{-4} \pm 7.12\times 10^{-7} $  &  $-4.05\times 10^{-5} \pm 5.23\times 10^{-7}$  &  $\pm 2.27\times 10^{-4}$    \\
 11  &  $-1.58\times 10^{-4} \pm 7.53\times 10^{-7}$  &  $-3.76\times 10^{-5} \pm 4.94\times 10^{-7}$  &  $\pm 2.31\times 10^{-4}$    \\
 12  &  $-1.60\times 10^{-4} \pm 8.01\times 10^{-7}$  &  $-3.18\times 10^{-5} \pm 4.48\times 10^{-7}$  &   $\pm 2.31\times 10^{-4}$    \\
 13  &  $-1.63\times 10^{-4} \pm 8.47\times 10^{-7}$  &  $-2.69\times 10^{-5} \pm 3.83\times 10^{-7}$  &  $\pm 2.37\times 10^{-4}$    \\
 14  &  $-1.66\times 10^{-4} \pm 8.96\times 10^{-7}$  &  $ -1.92\times 10^{-5} \pm 2.84\times 10^{-7}$  &   $\pm 2.40\times 10^{-4}$    \\
\end{tabular}
\end{ruledtabular}
\end{table}

\appendix

\section{Fermion propagator with chiral asymmetry}
\label{App:Propagator}

The fermion propagator with the nonzero Dirac mass, chiral shift, and chiral chemical potential is formally 
defined by 
\begin{equation} 
G(u,u') = i \langle u | \left[ ( i \partial_t + \mu ) \gamma^0 + \Delta \gamma^3 \gamma^5 + \mu_5 \gamma^0 \gamma^5 - (\bm{\pi}_{\perp} \cdot \bm{\gamma} ) 
 	    - \pi^3 \gamma^3 - m \right] ^{-1} | u' \rangle ,
\end{equation}
where $u=(t,\mathbf{r})$ is a space-time four-vector. By making use of this definition and utilizing the 
same method as in Ref.~\cite{chiral-shift-2}, we straightforwardly show that the fermion propagator 
has the general structure as in Eq.~(\ref{propagator-Phi-S}) and the explicit form of the translation 
invariant part is determined by the following expression of its Fourier transform:
\begin{equation}
\bar{G}(\omega,\mathbf{k}) = \int dt d^3\mathbf{r} e^{i\omega t-i(\mathbf{k}\cdot\mathbf{r})}
\bar{G} (t;\mathbf{r})
= i e^{-k_\perp^2l^2} \sum_{n=0}^{\infty} (-1)^n D_n (\omega,\mathbf{k}) \frac{1}{\mathcal{M} - 2n |eB| },
\label{G-omega-k}
\end{equation}
where $\mathbf{k}_\perp = (k^1, k^2)$, and the $n$th Landau level contribution is given in 
terms of the following matrix functions:
\begin{eqnarray}
D_n (\omega,\mathbf{k}) & = & 2 W \left[ \mathcal{P}_{-}L_n(2k_\perp^2l^2) - \mathcal{P}_{+}L_{n-1}(2k_\perp^2l^2) \right] + 4 (\mathbf{k}_\perp\cdot\bm{\gamma}) L^1_{n-1}(2k_\perp^2l^2),\\
W & = & ( \omega + \mu ) \gamma^0 - \Delta \gamma^3 \gamma^5 - \mu_5\gamma^0\gamma^5 - k^3\gamma^3 + m , \\
\mathcal{M} & = & ( \omega + \mu )^2 - \Delta^2 + \mu_5^2 - {k_3}^2 - m^2 - 2 [ \Delta k_3 + ( \omega + \mu ) \mu_5 ] \gamma^5 + 2m\Delta \gamma^3\gamma^5 + 2m\mu_5\gamma^0\gamma^5 .
\end{eqnarray}
Note that the last matrix factor in Eq.~(\ref{G-omega-k}) can be rewritten in the following equivalent form:
\begin{equation}
\frac{1}{\mathcal{M} - 2n |eB| }
 = \frac { \left[ ( \omega + \mu )^2 - \Delta^2 + \mu_5^2 - {k_3}^2 - m^2 - 2n|eB| \right] + 2 [ \Delta k_3 + ( \omega + \mu ) \mu_5 ] \gamma^5 - 2m\mu_5\gamma^0\gamma^5 - 2m\Delta \gamma^3\gamma^5 } 
              { \left[ ( \omega + \mu )^2 - \Delta^2 + \mu_5^2 - {k_3}^2 - m^2 - 2n|eB| \right]^2 - 4 [ \Delta k_3 + ( \omega + \mu ) \mu_5 ]^2 + 4m^2\mu_5^2 - 4 m^2 \Delta^2 },
\end{equation}
which implies that the fermion dispersion relation in the presence of the chiral asymmetry is determined by the solutions to the equation:
\begin{equation}
\left[ ( \omega + \mu )^2 - \Delta^2 + \mu_5^2 - {k_3}^2 - m^2 - 2n|eB| \right]^2 - 4 [ \Delta k_3 + ( \omega + \mu ) \mu_5 ]^2 + 4m^2\mu_5^2 - 4 m^2 \Delta^2 = 0 .
\end{equation}

\section{Chiral asymmetry functions in the HDL approximation}
\label{App:Chiral-asymmetry-functions}

In this appendix, we present the explicit form of the chiral asymmetry functions $\Delta_{n}(p_3)$ and $\mu_{5,n}(p_3)$ in 
the approximation with the HDL photon propagator. 

By making use of the definition in Eqs.~(\ref{Delta-n}) and (\ref{mu-5-n}), as well as the explicit form of the HDL photon 
propagator in Eq.~(\ref{D-HDL}), we derive the following results for the two coefficient functions of interest:
\begin{eqnarray}
\Delta_n(p_3)  & =  & (-1)^n e^2 l^2 \mathrm{sign}(eB) 
                        \int \frac{ dk_3 d^2 \mathbf{k}_{\perp}d^2\mathbf{p}_{\perp}}{(2\pi)^4} e^{-k_{\perp}^2l^2 - p^2_{\perp}l^2}  \displaystyle\sum_{N=0}^{\infty}  (-1)^N  \nonumber \\
&  &                       \Bigg\{ \left[  L_n(2p_{\perp}^2l^2) + L_{n-1}(2p_{\perp}^2l^2) \right] 
                        \left[ L_N(2k_{\perp}^2l^2) - L_{N-1}(2k_{\perp}^2l^2) \right] \left( - \mathcal{D}^{\rm (mag)} + \frac{1}{2} \mathcal{D}^{\rm (el)} \right) \nonumber \\ 
          &    &  +   \left[  L_n(2p_{\perp}^2l^2) - L_{n-1}(2p_{\perp}^2l^2) \right] 
                        \left[ L_N(2k_{\perp}^2l^2) + L_{N-1}(2k_{\perp}^2l^2) \right]  \left( \frac{ q_3^2 }{|\bm{q}|^{2}} \mathcal{D}^{\rm (mag)} + \frac{1}{2} \mathcal{D}^{\rm (el)} \right) \Bigg\} , 
\end{eqnarray}
\begin{eqnarray}
\mu_{5,n}(p_3)  & =  &  (-1)^n e^2 l^2 \mathrm{sign}(eB) 
                        \int \frac{ dk_3 d^2 \mathbf{k}_{\perp}d^2\mathbf{p}_{\perp}}{(2\pi)^4} e^{-k_{\perp}^2l^2 - p^2_{\perp}l^2}  \displaystyle\sum_{N=0}^{\infty} 
                         (-1)^N    \nonumber \\
&  &                       \Bigg\{ - \left[  L_n(2p_{\perp}^2l^2) + L_{n-1}(2p_{\perp}^2l^2) \right] \left[ L_N(2k_{\perp}^2l^2) - L_{N-1}(2k_{\perp}^2l^2) \right] k_3 \left( \frac{ q_3^2 }{|\bm{q}|^{2}} \mathcal{F}^{\rm (mag)} 
			+ \frac{1}{2} \mathcal{F}^{\rm (el)}  \right) \nonumber \\
         &    &    +   \left[  L_n(2p_{\perp}^2l^2) - L_{n-1}(2p_{\perp}^2l^2) \right] 
                        \left[ L_N(2k_{\perp}^2l^2) + L_{N-1}(2k_{\perp}^2l^2) \right] k_3 \left( \mathcal{F}^{\rm (mag)} - \frac{1}{2} \mathcal{F}^{\rm (el)} \right)  \nonumber \\
&  &               	+ 4  L^1_{N-1}(2k_{\perp}^2l^2) \left[  L_n(2p_{\perp}^2l^2) + L_{n-1}(2p_{\perp}^2l^2) \right] \frac{ q_3(k_1 q_1 + k_2 q_2 ) }{|\bm{q}|^{2}} \mathcal{F}^{\rm (mag)}  \Bigg\} ,
\end{eqnarray}
where the explicit expressions for the functions $\mathcal{D}^{\rm (mag)}$, $\mathcal{D}^{\rm (el)}$, $\mathcal{F}^{\rm (mag)}$, and 
$\mathcal{F}^{\rm (el)}$ are obtained after the integrations over $k_0\equiv i\omega_E$ performed, i.e.,  
\begin{eqnarray}
\mathcal{D}^{\rm (mag)}
& = & \frac{i}{\pi} |\bm{q}| \int_{-\infty} ^\infty \frac{ ( \omega_E - i \mu ) d\omega_E  }{ \left[ (\omega_E - i\mu )^2 + \mathcal{M}^2_N \right] \left( |\bm{q}|^3 + \frac{\pi}{4} m_D^2 |\omega_E + i p_0 | \right)  }\nonumber \\
& = &   \frac{  |\bm{q}|^4   \mathrm{sign}(\mu) \mathrm{sign}(\mathcal{M}_N^2 - \mu^2) }   { 2 \left[ |\bm{q}|^6 + (\frac{\pi}{4}m_D^2)^2 \left( \mathcal{M}_N-|\mu|  \right)^2 \right]} 
     - |\bm{q}| \mathrm{sign}(\mu)  \frac{ \frac{1}{4}m_D^2  \left(  \mathcal{M}_N- |\mu| \right)  \ln {\frac{|\bm{q}|^3}{ \frac{\pi}{4}m_D^2 | \mathcal{M}_N-|\mu|  | }   } } 
      {|\bm{q}|^6 + (\frac{\pi}{4}m_D^2)^2 \left( \mathcal{M}_N-|\mu|  \right)^2 }  \nonumber \\
& -&  \frac{  |\bm{q}|^4  \mathrm{sign}(\mu) }  { 2 \left[ |\bm{q}|^6 + (\frac{\pi}{4}m_D^2)^2 \left( \mathcal{M}_N+|\mu|  \right)^2 \right] }
            +  |\bm{q}|  \mathrm{sign}(\mu) \frac{ \frac{1}{4}m_D^2  \left(  \mathcal{M}_N+ |\mu| \right)  \ln {\frac{|\bm{q}|^3}{ \frac{\pi}{4}m_D^2 \left( \mathcal{M}_N+|\mu|  \right) }   } } 
      {|\bm{q}|^6 + (\frac{\pi}{4}m_D^2)^2 \left( \mathcal{M}_N+|\mu|  \right)^2 }  ,
\end{eqnarray}
\begin{eqnarray}
\mathcal{D}^{\rm (el)}
& = & \frac{1}{\pi} \int_{-\infty} ^\infty \frac{ ( i \omega_E + \mu ) d\omega_E  }{ \left[ (\omega_E - i\mu )^2 + \mathcal{M}^2_N \right] \left[ (\omega_E-ip_0)^2 + |\bm{q}|^2+m_D^2  \right]  } \nonumber \\
& = & \frac{\mu \, \theta[\mathcal{M}_N^2 - \mu^2 ] } {\sqrt{|\bm{q}|^2+m_D^2} \left[ \left( \sqrt{|\bm{q}|^2+m_D^2} + \mathcal{M}_N \right)^2 - \mu^2 \right] } 
       - \frac{ \mathrm{sign}(\mu) \theta \left[ \mu^2 - \mathcal{M}_N^2 \right] \left( \sqrt{|\bm{q}|^2+m_D^2} + |\mu| \right)  } {\sqrt{|\bm{q}|^2+m_D^2 }
           \left[ \left( \sqrt{|\bm{q}|^2+m_D^2} + |\mu| \right)^2 - \mathcal{M}_N^2  \right]} ,
\end{eqnarray}
and
\begin{eqnarray}
\mathcal{F}^{\rm (mag)}
& = & \frac{1}{\pi} |\bm{q}| \int_{-\infty} ^\infty \frac{ d\omega_E  }{ \left[ (\omega_E - i\mu )^2 + \mathcal{M}^2_N \right] \left( |\bm{q}|^3 + \frac{\pi}{4} m_D^2 |\omega_E + i p_0 | \right)  }\nonumber \\
& = &  \frac{1}{\mathcal{M}_N} \Bigg( \frac{ |\bm{q}|^4   \mathrm{sign}(\mathcal{M}_N^2 - \mu^2) }   {2 \left[ |\bm{q}|^6 + (\frac{\pi}{4}m_D^2)^2 \left( \mathcal{M}_N-|\mu|  \right)^2 \right] } 
     - |\bm{q}|  \frac{ \frac{1}{4}m_D^2  \left(  \mathcal{M}_N- |\mu| \right)  \ln {\frac{|\bm{q}|^3}{ \frac{\pi}{4}m_D^2 | \mathcal{M}_N-|\mu|  | }   } } 
      {|\bm{q}|^6 + (\frac{\pi}{4}m_D^2)^2 \left( \mathcal{M}_N-|\mu|  \right)^2 }  \nonumber \\
&    &   + \frac{ |\bm{q}|^4  } {2 \left[ |\bm{q}|^6 + (\frac{\pi}{4}m_D^2)^2 \left( \mathcal{M}_N+|\mu|  \right)^2 \right] }
            -  |\bm{q}|   \frac{ \frac{1}{4}m_D^2  \left(  \mathcal{M}_N+ |\mu| \right)  \ln {\frac{|\bm{q}|^3}{ \frac{\pi}{4}m_D^2 \left( \mathcal{M}_N+|\mu|  \right) }   } } 
      {|\bm{q}|^6 + (\frac{\pi}{4}m_D^2)^2 \left( \mathcal{M}_N+|\mu|  \right)^2 }  \Bigg) ,    
\end{eqnarray}
\begin{eqnarray}
\mathcal{F}^{\rm (el)}
& = & \frac{1}{\pi} \int_{-\infty} ^\infty \frac{  d\omega_E  }{ \left[ (\omega_E - i\mu )^2 + \mathcal{M}^2_N \right] \left[ (\omega_E-ip_0)^2 + |\bm{q}|^2+m_D^2  \right]  } \nonumber \\
& = & - \frac{ \theta[ \mu^2 - \mathcal{M}_N^2 ] } {\sqrt{|\bm{q}|^2+m_D^2} \left[ \left( \sqrt{|\bm{q}|^2+m_D^2} + |\mu| \right)^2 - \mathcal{M}_N^2 \right] } 
       + \frac{ \theta \left[ \mathcal{M}_N^2 -\mu^2 \right] \left( \sqrt{|\bm{q}|^2+m_D^2} + \mathcal{M}_N \right)  } { \sqrt{|\bm{q}|^2+m_D^2 } \mathcal{M}_N
           \left[ \left( \sqrt{|\bm{q}|^2+m_D^2} + \mathcal{M}_N \right)^2 - \mu^2  \right]}  ,
\end{eqnarray}
where 
$\mathcal{M}_N^2 = k_3^2 + 2N|eB| + m^2$ and 
$|\bm{q}|^2 = | k_3 - p_3 |^2 + k_\perp^2 + p_\perp^2 - 2 k_\perp p_\perp \cos \phi$.

While performing the numerical analysis, it is convenient to render the above expressions in a dimensionless
form. Therefore, we introduce the following dimensionless functions: $\bar{\Delta}_n \equiv  \Delta_n/\mu$ and 
$\bar{\mu}_{5,n} \equiv  \mu_{5,n}/\mu$, as well as the following dimensionless variables: 
$x = p_{\perp}/\mu$, $y \equiv  k_{\perp}/\mu$,  $ x_{3} \equiv p_{3}/\mu$, and $y_{3} \equiv k_{3}/\mu$.
By using this new notation, we have 
\begin{eqnarray}
\bar{\Delta}_n  & =  & (-1)^n \frac {e^2}{b} \mathrm{sign}(eB) 
                        \int \frac{ dy_3 dy dx d\phi}{(2\pi)^3} e^{-(x^2+y^2)/b }  \displaystyle\sum_{N=0}^{\infty}  (-1)^N xy  \nonumber \\
& \times &                       \Bigg\{ \left[  L_n(2x^2/b) + L_{n-1}(2x^2/b) \right] 
                        \left[ L_N(2y^2/b) - L_{N-1}(2y^2/b) \right] \left( \mathcal{ - \bar{D}}^{\rm (mag)} + \frac{1}{2} \mathcal{\bar{D}}^{\rm (el)} \right) \nonumber \\ 
          &  -  &    \left[  L_n(2x^2/b) - L_{n-1}(2x^2/b) \right] 
                        \left[ L_N(2y^2/b) + L_{N-1}(2y^2/b) \right]  \left( \frac{ (x_3-y_3)^2 \mathcal{\bar{D}}^{\rm (mag)} }{(x_3-y_3)^2+x^2+y^2-2xy\cos\phi}  + \frac{1}{2} \mathcal{\bar{D}}^{\rm (el)} \right) \Bigg\} ,
\label{Delta-n-dimless}
\end{eqnarray}
and 
\begin{eqnarray}
\bar{\mu}_{5,n}  & =  & (-1)^n \frac {e^2}{b} \mathrm{sign}(eB) 
                        \int \frac{ dy_3 dy dx d\phi}{(2\pi)^3} e^{-(x^2+y^2)/b }  \displaystyle\sum_{N=0}^{\infty}  (-1)^N xy y_3 \nonumber \\
& \times &                       \Bigg\{ - \left[  L_n(2x^2/b) + L_{n-1}(2x^2/b) \right] 
                        \left[ L_N(2y^2/b) - L_{N-1}(2y^2/b) \right]  \left( \frac{ (x_3-y_3)^2 \mathcal{ \bar{F}}^{\rm (mag)}  }{(x_3-y_3)^2+x^2+y^2-2xy\cos\phi} + \frac{1}{2} \mathcal{\bar{F}}^{\rm (el)} \right) \nonumber \\ 
          &  +  &  \left[  L_n(2x^2/b) - L_{n-1}(2x^2/b) \right] 
                        \left[ L_N(2y^2/b) + L_{N-1}(2y^2/b) \right]  \left( \mathcal{\bar{F}}^{\rm (mag)} - \frac{1}{2} \mathcal{\bar{F}}^{\rm (el)} \right) \Bigg\} ,
\label{mu-5-n-dimless}
\end{eqnarray}
where 
\begin{eqnarray}
\bar{\mathcal{D}}^{\rm (mag)}
& = &   \frac{   |\bm{\bar{q}}|^4   \mathrm{sign}(\mu) \mathrm{sign}(\mathcal{\bar{M}}_N^2 - 1) }   {2 \left[ |\bar{\bm{q}}|^6 + (\frac{\pi d^2}{4})^2 \left( \mathcal{\bar{M}}_N- \mathrm{sign}(\mu)  \right)^2 \right] } 
     - |\bm{\bar{q}}| \mathrm{sign}(\mu)  \frac{ \frac{d^2}{4}  \left(  \mathcal{\bar{M}}_N- \mathrm{sign}(\mu) \right)  \ln {\frac{|\bm{\bar{q}}|^3}{ \frac{\pi d^2}{4} | \mathcal{\bar{M}}_N- \mathrm{sign}(\mu) | }   } } 
      {|\bm{\bar{q}}|^6 + (\frac{\pi d^2}{4})^2 \left( \mathcal{M}_N-\mathrm{sign}(\mu)  \right)^2 }  \nonumber \\
&    &   - \frac{   |\bm{\bar{q}}|^4  \mathrm{sign}(\mu) }  { 2 \left[ |\bar{\bm{q}}|^6 + (\frac{\pi d^2}{4})^2 \left( \mathcal{\bar{M}}_N+\mathrm{sign}(\mu)  \right)^2 \right] }
            +  |\bm{\bar{q}}|  \mathrm{sign}(\mu) \frac{ \frac{d^2}{4}  \left(  \mathcal{\bar{M}}_N+ \mathrm{sign}(\mu) \right)  \ln {\frac{|\bm{\bar{q}}|^3}{ \frac{\pi d^2}{4} \left( \mathcal{\bar{M}}_N+\mathrm{sign}(\mu) \right) }   } } 
      {|\bm{\bar{q}}|^6 + (\frac{\pi d^2}{4})^2 \left( \mathcal{\bar{M}}_N+\mathrm{sign}(\mu)  \right)^2 }     ,
\end{eqnarray}
\begin{eqnarray}
\bar{\mathcal{D}}^{\rm (el)}
& = & \frac{ \theta[\mathcal{\bar{M}}_N^2 - 1 ] } {\sqrt{|\bar{\bm{q}}|^2+d^2} \left[ \left( \sqrt{|\bar{\bm{q}}|^2+d^2} + \mathcal{\bar{M}}_N \right)^2 - 1 \right] } 
       - \frac{ \mathrm{sign}(\mu) \theta \left[ 1 - \mathcal{\bar{M}}_N^2 \right] \left( \sqrt{|\bar{\bm{q}}|^2+d^2} + \mathrm{sign}(\mu) \right)  } {\sqrt{|\bar{\bm{q}}|^2+d^2 }
           \left[ \left( \sqrt{|\bar{\bm{q}}|^2+d^2} + \mathrm{sign}(\mu) \right)^2 - \mathcal{\bar{M}}_N^2  \right]} ,
\end{eqnarray}
\begin{eqnarray}
\bar{\mathcal{F}}^{\rm (mag)}
& = &   \frac{1}{\mathcal{\bar{M}}} \Bigg\{ \frac{   |\bm{\bar{q}}|^4   \mathrm{sign}(\mathcal{\bar{M}}_N^2 - 1) }  
        {2 \left[ |\bar{\bm{q}}|^6 + (\frac{\pi d^2}{4})^2 \left( \mathcal{\bar{M}}_N- \mathrm{sign}(\mu)  \right)^2 \right] } 
     - |\bm{\bar{q}}|  \frac{ \frac{d^2}{4}  \left(  \mathcal{\bar{M}}_N- \mathrm{sign}(\mu) \right)  \ln {\frac{|\bm{\bar{q}}|^3}{ \frac{\pi d^2}{4} | \mathcal{\bar{M}}_N- \mathrm{sign}(\mu) | }   } } 
      {|\bm{\bar{q}}|^6 + (\frac{\pi d^2}{4})^2 \left( \mathcal{M}_N-\mathrm{sign}(\mu)  \right)^2 }  \nonumber \\
&    &   + \frac{   |\bm{\bar{q}}|^4  }  { 2 \left[ |\bar{\bm{q}}|^6 + (\frac{\pi d^2}{4})^2 \left( \mathcal{\bar{M}}_N+\mathrm{sign}(\mu)  \right)^2 \right] }
            -  |\bm{\bar{q}}|  \frac{ \frac{d^2}{4}  \left(  \mathcal{\bar{M}}_N+ \mathrm{sign}(\mu) \right)  \ln {\frac{|\bm{\bar{q}}|^3}{ \frac{\pi d^2}{4} \left( \mathcal{\bar{M}}_N+\mathrm{sign}(\mu) \right) }   } } 
      {|\bm{\bar{q}}|^6 + (\frac{\pi d^2}{4})^2 \left( \mathcal{\bar{M}}_N+\mathrm{sign}(\mu)  \right)^2 } \Bigg\} ,       
\end{eqnarray}
\begin{eqnarray}
\bar{\mathcal{F}}^{\rm (el)}
& = & - \frac{ \theta[ 1 - \mathcal{ \bar{M}}_N^2 ]} {\sqrt{|\bar{\bm{q}}|^2+d^2} \left[ \left( \sqrt{|\bar{\bm{q}}|^2+d^2} + 1 \right)^2 - \mathcal{\bar{M}}_N^2 \right] } 
    + \frac{ \theta \left[ \mathcal{\bar{M}}_N^2 - 1 \right] \left( \sqrt{|\bar{\bm{q}}|^2+d^2} + \mathcal{\bar{M}}_N \right)  } {\mathcal{\bar{M}}_N \sqrt{|\bar{\bm{q}}|^2+d^2 }
           \left[ \left( \sqrt{|\bar{\bm{q}}|^2+d^2} +\mathcal{\bar{M}}_N \right)^2 - 1 \right]} ,
\end{eqnarray}
with $\mathcal{\bar{M}}_N^2 = y_3^2 + 2Nb + a^2$ and $|\bar{\bm{q}}|^2 = ( y_3 - x_3 )^2 + y^2 + x^2 - 2 xy \cos \phi$.
Note that the dimensionless parameters $a$, $b$, and $d$ are defined in Eqs.~(\ref{def-a}), (\ref{def-b}), and
(\ref{def-d}), respectively.

It is instructive to note that the function under the integral in the expression for $\bar{\mu}_{5,n}$ contains 
an overall factor of $y_3$ in the numerator. Clearly, such a dependence on $y_3$ is not very helpful for 
the numerical convergence of the integral. By taking into account, however, that the rest of the integrand 
depends on $y_3$ only via $( y_3 - x_3 )^2$ and $y_3^2$ combinations, the convergence can be 
substantially improved by using the following identity:
\begin{equation}
\int_{-\infty}^{\infty} dy_3 \, y_3 \, F\left((y_3-x_3)^2, y_3^2 \right) 
= \int_{-\infty}^{\infty}  dy_3  \, \frac{y_3}{2}  \left[ F\left((y_3-x_3)^2, y_3^2 \right) - F\left((y_3+x_3)^2, y_3^2 \right) \right].
\end{equation}

\end{document}